\documentclass[final,2p,times,twocolumn]{elsarticle}

\usepackage{graphicx}
\usepackage{epstopdf}

\usepackage{amssymb}

\newcommand{\lacls}{LaCl$_{3}$(Ce) }

\journal{Nucl. Instr. and Meth. in Phys. Res. A}
\begin{document}

\begin{frontmatter}

\title{First i-TED demonstrator: a Compton imager with Dynamic Electronic Collimation}

\author{V.~Babiano, J. Balibrea, L.~Caballero, D.~Calvo, I.~Ladarescu, J.~Lerendegui, S.~Mira Prats, C.~Domingo-Pardo\footnote{Corresponding author.}}
\address{Instituto de F{\'\i}sica Corpuscular, CSIC-University of Valencia, Spain}


\begin{abstract}
  i-TED consists of both a total energy detector and a Compton camera primarily intended for the measurement of neutron capture cross sections by means of the simultaneous combination of neutron time-of-flight (TOF) and $\gamma$-ray imaging techniques. TOF allows one to obtain a neutron-energy differential capture yield, whereas the imaging capability is intended for the discrimination of radiative background sources, that have a spatial origin different from that of the capture sample under investigation. A distinctive feature of i-TED is the embedded Dynamic Electronic Collimation (DEC) concept, which allows for a trade-off between efficiency and image resolution. Here we report on some general design considerations and first performance characterization measurements made with an i-TED demonstrator in order to explore its $\gamma$-ray detection and imaging capabilities.
  
\end{abstract}


\end{frontmatter}

\section{Introduction}\label{sec:introduction}

Neutron capture cross sections are a key nuclear-physics ingredient for the study of the formation of elements heavier than iron in the stars~\cite{BBFH,Cameron57,Kaeppeler11}. Direct ($n,\gamma$) measurements are particularly relevant for the so-called slow nucleosynthesis ($s$-) process, where most of the involved nuclei can be accessed in laboratory experiments. Presently, the main fronts of experimental research in this field focus, on one hand, on new measurements targeting accuracies of $\lesssim$5\% required by the models~\cite{Nishimura17,Cescutti18} and, on the other hand, on the first measurement of radioactive $s$-process branching nuclei, which are particularly relevant for benchmarking the physical conditions inside stars~\cite{Kaeppeler11}.

Neutron activation and time-of-flight (TOF) are the two main methodologies used to determine neutron capture cross sections in the relevant stellar energy range (see e.g.~\cite{Kaeppeler11}). The TOF approach offers the advantage of being a general technique applicable to any isotope, provided that sufficient sample mass becomes available. It also allows one to cover the full stellar energy regime in one single measurement. For TOF experiments essentially two different $\gamma$-ray detection apparatus are utilized: Total Absorption Calorimeters (TACs)~\cite{Wisshak89,Reifarth05,Guerrero12a} and Total-Energy Detectors (TEDs)~\cite{Macklin67,Abbondanno04,Borella07}. TEDs have been used since many years for neutron capture cross section measurements of astrophysical interest, including experiments with radioactive samples of $s$-process branching nuclei~\cite{Winters87,Abbondanno04b,Lederer13}. However, one limitation of state-of-the-art TEDs~\cite{Plag03} arises from neutrons scattered in the capture-sample, and subsequently thermalized and captured in the surrounding walls and structural materials in the measuring station. This effect has been investigated in detail in Ref.~\cite{Zugec14} and, in some important cases (see e.g.~\cite{Tagliente13,Neyskens15}), it fully dominates the background level already beyond a few keV of neutron energy. This is precisely the energy range of interest for nucleosynthesis studies in low-mass AGB stars during H-burning (8~keV) and He-burning (23~keV), and also during core He-burning (26~keV) and shell C-burning (91~keV) in massive stars~\cite{Kaeppeler11}. Heavy lead shieldings (see e.g. Fig.~3 in Ref.~\cite{Walter86}) or mechanical collimators~\cite{Perez16} become prohibitive owing to the increased neutron scattering and capture in the shielding itself, which rather amplifies the problem of the neutron-induced $\gamma$-ray background.

One alternative to this problem is to use electronic collimation~\cite{Schoenfelder73,Everett77}, which is based on the Compton scattering law. In the framework of the HYMNS project~\cite{hymns} we are developing an electronically collimated TED to be applied for neutron capture TOF measurements. As described in Ref.~\cite{Domingo16} this approach would allow one, on an event-by-event basis, to infer information on the incoming radiation direction, thereby enabling the possibility to distinguish between true capture $\gamma$-rays coming from the sample and contaminant $\gamma$-rays induced by neutron-capture in the surroundings of the detection set-up. Similar concepts have been exploited in other fields, such as $\gamma$-ray astronomy~\cite{Schoenfelder04,Ichinohe16,Tajima18}, medicine~\cite{Aldawood17} or nuclear structure~\cite{Tashenov08} but thus far, it has never been attempted in the field of TOF neutron-capture measurements.
In this work we present a detection system, which represents a first demonstrator of the so-called Total-Energy Detector with $\gamma$-ray imaging capability (i-TED)~\cite{Domingo16}.  The suitability of i-TED as a total-energy detector has been discussed on the basis of MC simulations in Ref.~\cite{Domingo16} and an experimental study using a neutron beam will be reported in a separate article. In this work we focus on the suitability of i-TED as a $\gamma$-ray imager. General considerations related to the Compton camera design are discussed in Sec.~\ref{sec:methods}. Sec.~\ref{sec:materials} describes the technical aspects of the first assembled i-TED demonstrator. Measurements made in the laboratory to characterize the efficiency and resolution are discussed in Sec.~\ref{sec:results}. Finally, Sec.~\ref{sec:summary} summarizes the main conclusions and next steps.

\section{Dynamic Electronic Collimation}\label{sec:methods}

Detection efficiency and image resolution are two of the most relevant performance aspects of any Compton imager. These two elements are discussed in the following, with the aim of introducing the concept of Dynamic Electronic Collimation (DEC).

In Compton cameras based on two detection planes the intrinsic $\gamma$-ray detection efficiency is determined by the efficiency of its scatter ($S$) and absorber ($A$) detectors and by the distance between them, hereafter indicated by $d_f$. The subindex $f$ is used here to reflect that the $S$-$A$ separation length, $d_f$, may be used to ``focus'' or tune the system, and thus adjust the interplay between image resolution and efficiency, as it is discussed below.

\begin{figure}[htbp!]
\flushleft
\centering
\includegraphics[width=0.46\columnwidth]{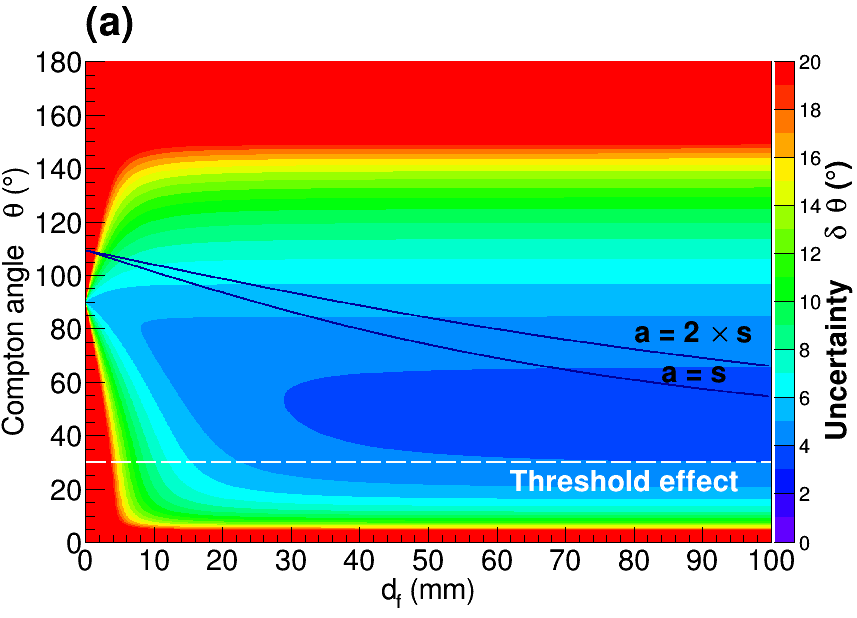}
\includegraphics[width=0.46\columnwidth]{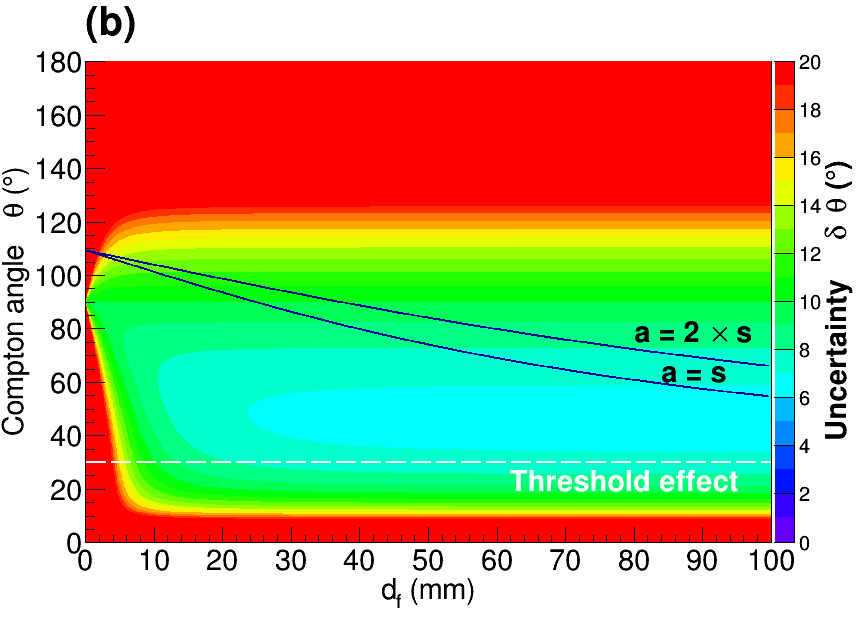}
\includegraphics[width=0.46\columnwidth]{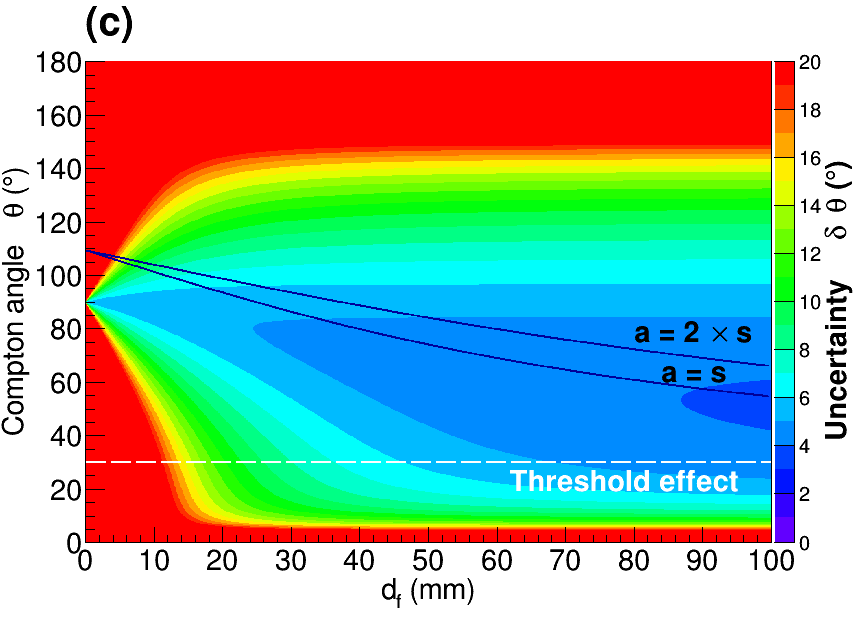}
\includegraphics[width=0.46\columnwidth]{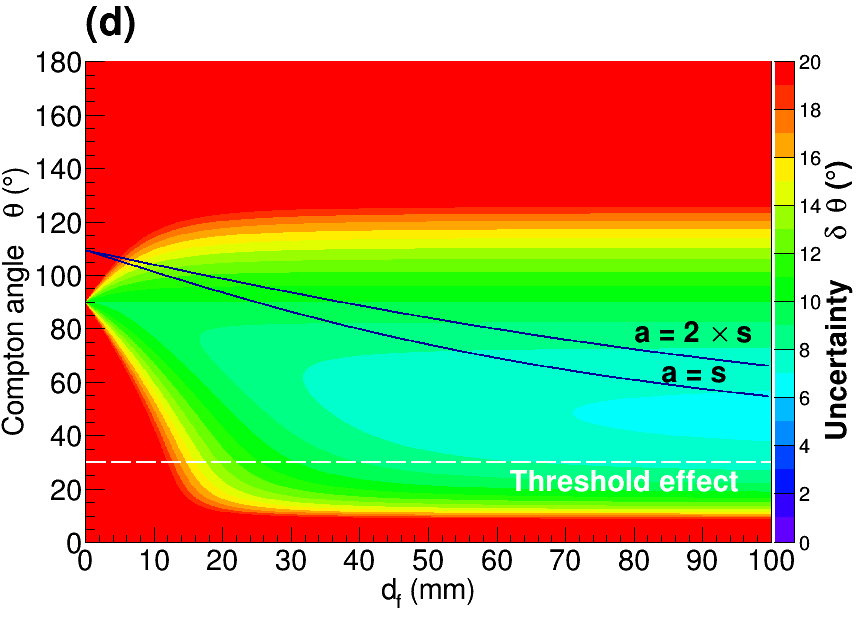}
\caption{\label{fig:2dresolution} Angular resolution calculated for a Compton camera as a function of the distance $d_f$ between scatter and absorber detectors. From top to bottom the spatial and energy resolutions (\textsc{fwhm}) are of (a) 1~mm, 3.5\%, (b) 1~mm, 6.5\%, (c) 3~mm, 3.5\% and (d) 3 mm, 6.5\%. See text for details.}
\end{figure}

Assuming that the energy $E$ of the incident $\gamma$-ray is known, the uncertainty on the Compton angle $\delta \theta$ can be approximated by the following analytic expression (see Ref.~\cite{Mihailescu07}),

\begin{equation}\label{eq:resolution}
\delta \theta = \frac{1}{\sin \theta} \left( \left( \frac{m_e c^2}{E'^2}\right)^2\left(\frac{\delta
  E'}{E'}\right)^2 + 2 \sin^2 \theta \left(\frac{\delta r}{r}\right)^2 \right)^{1/2},
\end{equation}
where $\theta$ is the Compton angle, $m_e$ is the rest mass of the electron, $c$ is the speed of light, $r$ is the distance between the first and the second interactions, and $E'$ is the energy of the scattered $\gamma$-ray.  Uncertainties on these quantities are indicated by the prefix $\delta$. Obviously, the first parameters to optimize in a Compton imager are the energy-resolution $\delta E$ and the spatial-resolution $\delta r$, because low uncertainties on these magnitudes lead to high angular resolution and imaging capabilities. In order to gain insight into this aspect, let us assume that the $\gamma$-ray source is sufficiently far from the detector, so that the distance between the first and the second interactions can be approximated by $r \simeq d_f/\cos(\theta)$. With such approximation one can illustrate in a simple manner the overall trend of the Compton angle uncertainty, or at least an upper limit for it. Thus, the uncertainty on the Compton angle $\delta \theta$ can be graphically displayed as a function of the $S$-$A$ separation $d_f$, as demonstrated in Fig.~\ref{fig:2dresolution} for a $\gamma$-ray source of 662~keV.
The uncertainty $\delta \theta$ is shown in color scale for a series of spatial resolutions $\delta r$ = 1~mm~\cite{Li10,Babiano19}, 3~mm~\cite{Gostojic16,Liprandi17,Ulyanov17,Babiano19} \textsc{fwhm} and energy-resolutions $\delta E'/E'$ = 3.5\%, 6.5\% \textsc{fwhm} at 662~keV, representative of high-resolution inorganic scintillation crystals (see Ref.~\cite{Olleros18} and references therein). For the graphical representation the value of $E'$ in eq.~(\ref{eq:resolution}) was calculated using the Compton law for each scattering angle $\theta$. For the relative energy resolution $\delta E'/E'$ a $1/\sqrt E$ dependence was assumed with respect to the values quoted at 662~keV. The Doppler contribution to $\delta E'$ due to the momentum of the target electron~\cite{Ribberfors75} can be fully neglected because it is much smaller than the intrinsic energy resolution of the inorganic crystals used in this work.

Regarding the quoted resolutions on $\delta E$ and $\delta r$, clearly, a good energy resolution (3.5\% instead of 6.5\%) has a large impact on the attainable angular (image) resolution, as it can be appreciated comparing Fig.\ref{fig:2dresolution} (a) and (c) with  Fig.\ref{fig:2dresolution} (b) and (d). On the other hand, the intrinsic spatial resolution of the radiation detectors has a lesser impact on the attainable angular resolution $\delta \theta$. This effect can be observed by the small changes in $\delta \theta$ when the spatial resolution varies from 1~mm Fig.\ref{fig:2dresolution} (a) and (b) to 3~mm Fig.\ref{fig:2dresolution} (c) and (d). The latter statement is particularly clear for sufficiently large $S$-$A$ separation lengths of $d_f \gtrsim 30$~mm.

\begin{figure}[htbp!]
\flushleft
\centering
\includegraphics[width=0.75\columnwidth]{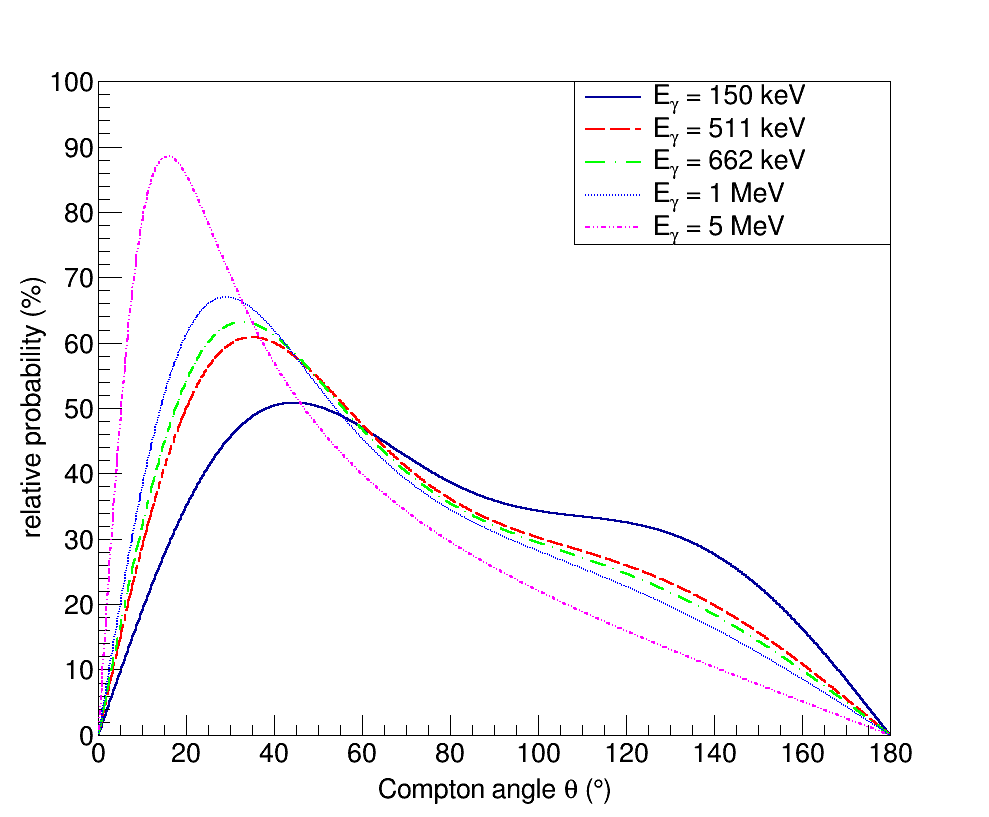}
\caption{\label{fig:KN} Relative scattering probability as a function of the Compton angle for a series of $\gamma$-ray energies between 150~keV and 5~MeV.}
\end{figure}

For $\gamma$-ray energies different from 662~keV the general pattern of Fig.~\ref{fig:2dresolution} remains rather similar, with the best (minimum) value of $\delta \theta$ changing from Compton angles of $\theta \sim 45^{\circ}$ at 662~keV (Fig.~\ref{fig:2dresolution}) down to $\theta \sim 40^{\circ}$ ($20^{\circ}$) at 1~MeV (5~MeV), and to $\theta \sim 70^{\circ}$ at 150~keV. Thus, for each incident $\gamma$-ray energy there is a small shift between the Compton angle for the best angular resolution and the Compton angle corresponding to the maximum probability of the Klein-Nishina distribution~\cite{Klein29} (see Fig.~\ref{fig:KN}). Nevertheless, both follow the same trend with the $\gamma$-ray energy, which is a positive feature in terms of image reconstruction.

In general, Compton angles around 0$^{\circ}$ and 180$^{\circ}$ lend to poor angular resolutions owing to the $1/\sin(\theta)$ factor in eq.~(\ref{eq:resolution}).  The contribution of the first term in eq.~(\ref{eq:resolution}) increases with the inverse value of the scattered $\gamma$-ray energy $E'$ to the fourth power. Hence, due to the Compton scattering law and the energy dependency of the resolution, the angular resolution worsens significantly above $\theta \sim 140^{\circ}$, or even $\sim 120^{\circ}$, depending on the intrinsic energy resolution of the detector (see Fig.~\ref{fig:2dresolution}).

On the other hand, the low range of scattering angles $10^{\circ} \lesssim \theta \lesssim 120^{\circ}(140^{\circ})$ represents the most valuable contribution in terms of image reconstruction (see Fig.~\ref{fig:2dresolution}). In this high-resolution angular range ($10^{\circ} \lesssim \theta \lesssim 120^{\circ}$) the second term in eq.~(\ref{eq:resolution}) contributes most, thus allowing one to enhance the angular resolution by means of increasing the $S$-$A$ separation or $d_f$ length. Because there is also a strong dependency between $\gamma$-ray efficiency and $d_f$ distance, the latter can be adjusted for a trade-off between detection efficiency and image resolution, or for maximizing either of them. It is also worth noting that, for the quoted intrinsic position- and energy-resolutions, the maximal useful $S$-$A$ separation $d_f$ is of about 40-60~mm in all cases. Indeed, beyond that distance there is essentially no gain in resolution and only a strong decrease in efficiency would be obtained.

The concept of Dynamic Electronic Collimation (DEC)~\cite{PCT16} enhances the fieldability and applicability scope of the Compton camera, when compared to a system where both $S$- and $A$-detection planes are placed at a fixed distance. Several previous studies have reported on the impact that the $S$-$A$ distance has on angular resolution~\cite{Boston07,Kataoka13} and on both angular resolution and efficiency~\cite{Takeuchi14,Lee15,Kishimoto17,Munoz17}, thereby focusing mainly on optimizing the detector geometry. More recently, the suitability of a Compton system with manually adjustable $S$-$A$ separation has been demonstrated in the field of 3D molecular Compton imaging by the group of Tokyo~\cite{Kishimoto17,Kishimoto17b}. With the aim of taking full advantage of the DEC concept, the i-TED Compton imager presented in this work incorporates an embedded micro-positioning drive with a large range of 50~mm, which enables for a variable and accurate separation of the $S$-$A$ detectors distance $d_f$. More technical details are given below in Sec.~\ref{sec:materials}. 

The main advantage of DEC for the measurement of neutron capture cross sections is envisaged for the reduction of neutron-induced background radiation. Indeed, in neutron TOF measurements the signal-to-background ratio typically decreases with increasing neutron energy due to the $1/v$ dependency of the cross section and the relatively flat background rate (see for example Fig.~6 in Ref.~\cite{Zugec14}). Thus, it may become possible to sample a broad neutron energy range using different values of $d_f$, which are adjusted to each specific neutron-energy decade or energy interval for an optimal background discrimination and sufficient counting statistics.

The DEC-technique may be also of interest in other fields such as homeland security scenarios, decommissioning and nuclear waste monitoring. Given the variety of $\gamma$-ray emitters, activities and spatial distributions, a first inspection using a short $d_f$ may be useful for a rapid survey to roughly identifiy the main radionuclides and their coarse spatial origin. Once this information has been processed, if needed, a more detailed image can be obtained aftewards using a larger $d_f$ value. DEC will be further discussed in Sec.~\ref{sec:results} on the basis of dedicated laboratory measurements. 

Finally, there are two experimental effects which constrain remarkably the angular resolution displayed in Fig.~\ref{fig:2dresolution}. They are discussed in the following two sections.

\subsection*{Geometry constraints}
\begin{figure}[htbp!]
\flushleft
\centering
\includegraphics[width=0.95\columnwidth]{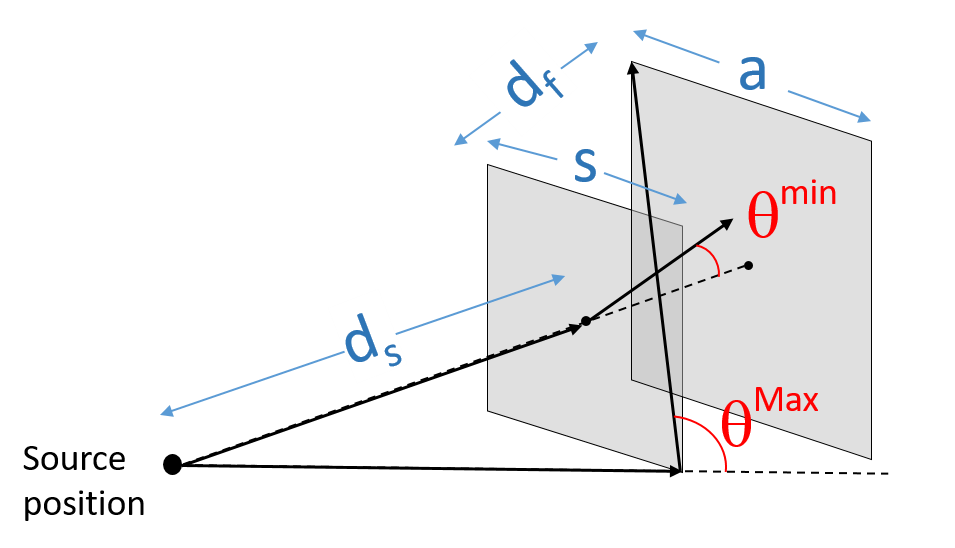}
\caption{\label{fig:sketch} Schematic representation of a generic Compton camera. The maximum and minimum Compton angles are represented for a $\gamma$-ray source centered at a distance $d_s$ from the Camera. See text for details.}
\end{figure}

The geometry of the detection setup determines to a large extent the total detection efficiency, but it also constrains the maximum Compton angle that can be measured. This is schematically illustrated in Fig.~\ref{fig:sketch} for a simplified model based on two parallel aligned square-shaped detectors.
Assuming $S$- and $A$-detectors of sizes $s$ and $a$, respectively, the maximum measurable Compton angle for a centered source at a distance $d_{s}$ is given by,
\begin{equation}\label{eq:geometry}
  \theta^{Max} = \arctan \left( \frac{s}{\sqrt 2 \cdot d_s} \right) + \arctan \left(  \frac{a+s}{\sqrt 2 \cdot d_f} \right).
\end{equation}

\begin{figure}[htbp!]
\flushleft
\centering
\includegraphics[width=\columnwidth]{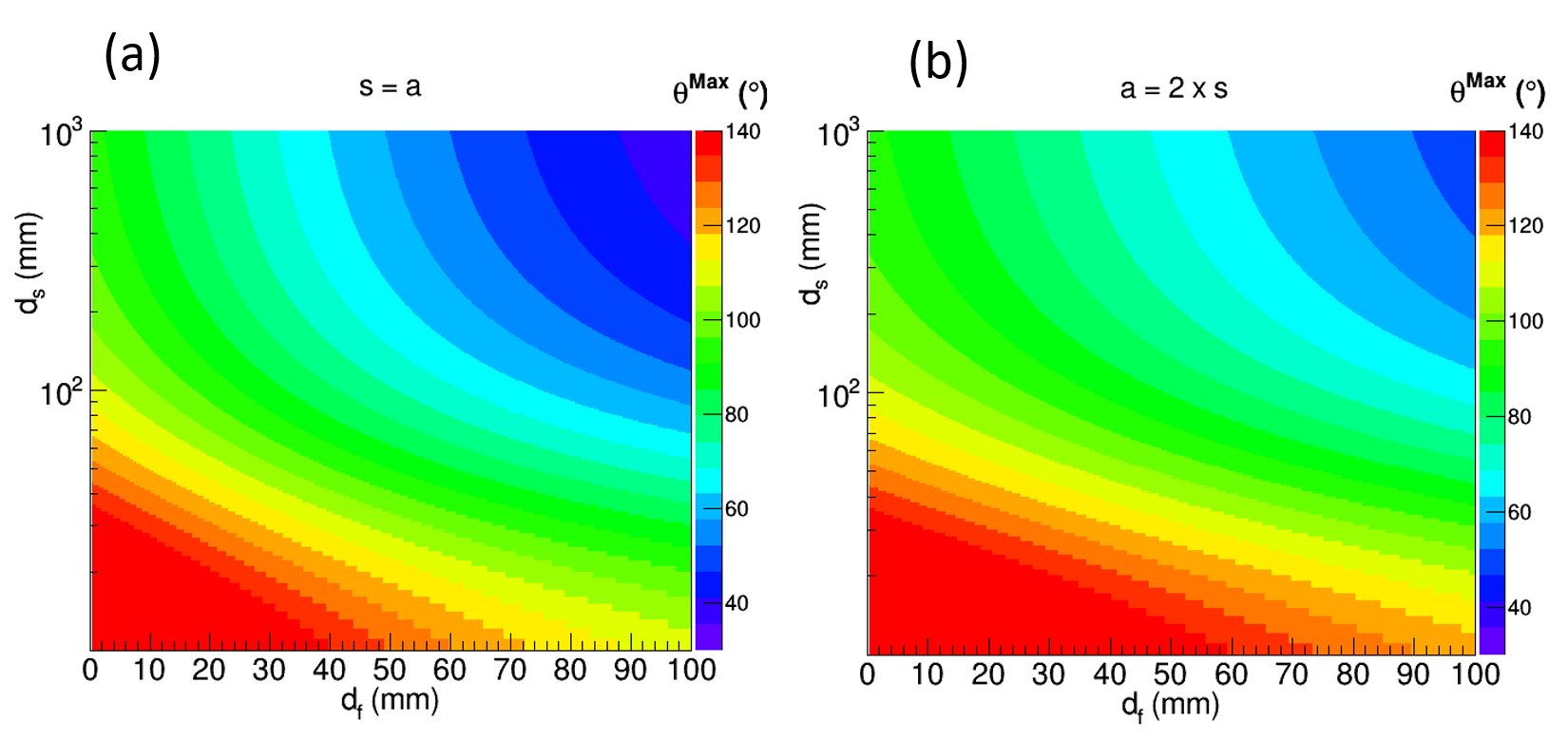}
\caption{\label{fig:max_angle} Representation of the maximum measurable Compton angle $\theta^{Max}$ with a color scale (in degrees) for a generic Compton camera based on squared detectors arranged in parallel configuration. Figure (a) shows the behavior for a camera based on S- and A-detectors of equal size ($s=a=50$~mm). Figure (b) corresponds to a configuration where the absorber detector has two times the size of the scatter detector ($a=2\times s = 100$~mm). See text for details.}
\end{figure}

Note that in the latter equation, both terms cannot be grouped in one single term using the arctangent addition formula $\arctan(u) \pm \arctan(v) = \arctan((u \pm v)/(1\mp u \cdot v))$ because the latter holds only for $u \cdot v < 1$. The value of $\theta^{Max}$ is displayed in Fig.~\ref{fig:max_angle} in color scale as a function of $d_f$ and the distance to a centered point-like source $d_s$. Two different cases are considered here. Fig.~\ref{fig:max_angle}-(a) shows the expected behavior for a Compton camera where both S- and A-detectors are squared and have the same size $a=s=50$~mm. Fig.~\ref{fig:max_angle}-(b) shows the maximum measurable Compton angle when the size of the A-detector is two times the size of the S-detector ($a=2\times s=100$~mm).

For a $\gamma$-ray source located at a short distance ($d_s \sim cm$) practically all relevant (high resolution) Compton angles are available over the main range of $S$-$A$ separation lengths $d_f = 10-60$~mm. For sources at distances of $d_s \sim 10$~cm, the maximum measurable Compton angle is constrained to $\theta^{Max}\lesssim$80$^{\circ}$. This constraint represents already a significant loss in terms of usable Compton angles. Such $\theta^{Max}$ limit for a source at $d_s = 10$~cm is shown in Fig.~\ref{fig:2dresolution} for the $s=a$ and $a=2\times s$ configurations as a function of $d_f$. It is worth noting that this is a purely geometrical constraint, which does not depend on the $\gamma$-ray energy. For the location of distant sources, $d_s \gtrsim 1$~m, using a $d_f=50$~mm the maximum detectable Compton angle $\theta^{Max}$ becomes $\sim 60^{\circ}$ and $\sim 80^{\circ}$ for the $s=a$ and $a = 2\times s$ configurations, respectively. Thus, a large absorber plane seems particularly interesting for the measurement of low energy distant sources (see also Fig.~\ref{fig:KN}). A significant portion of usable Compton angles are available yet below $\theta^{Max} = 60^{\circ}-80^{\circ}$ over a broad range of energies (see Fig.~\ref{fig:KN}). However, the low angular range is constrained by another experimental effect, which is described below.

\subsection*{Threshold effect}
\begin{figure}[htbp!]
\flushleft
\centering
\includegraphics[width=\columnwidth]{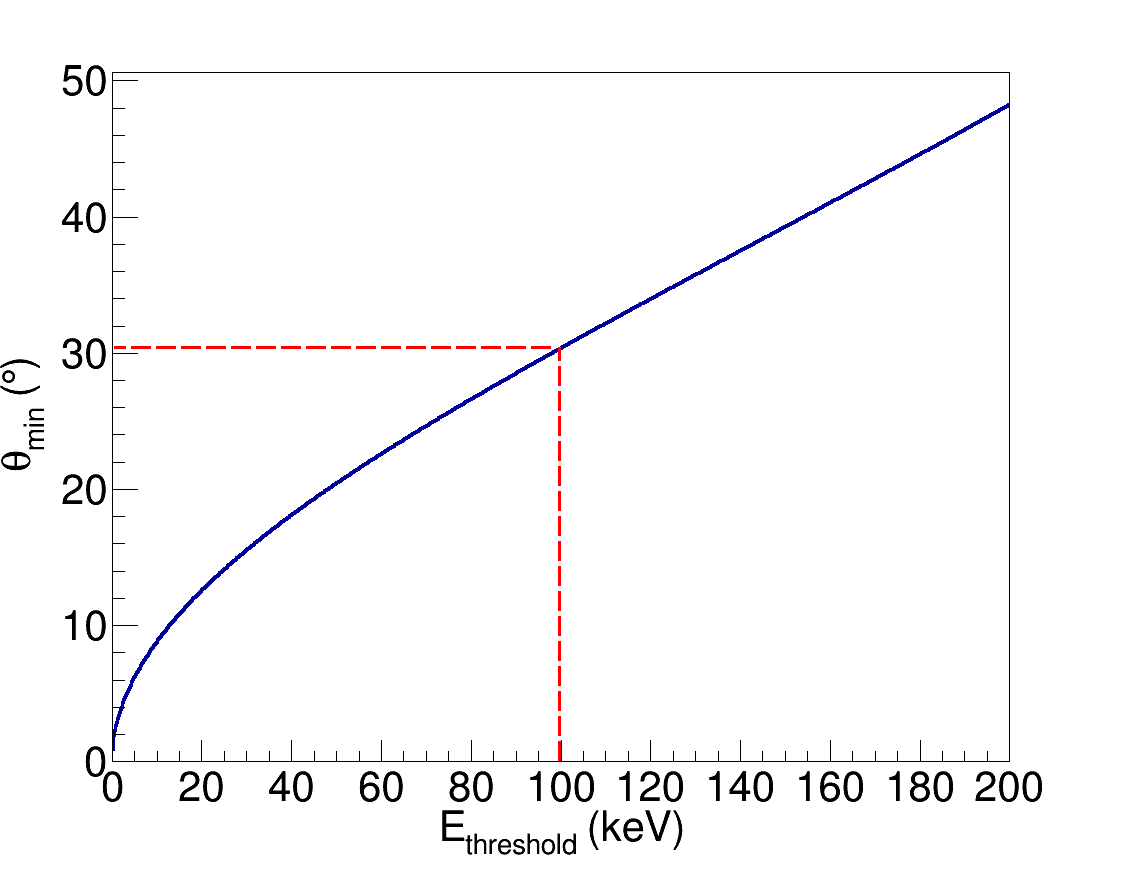}
\caption{\label{fig:threshold} Relationship between the energy threshold in the scatter detector and the minimum measurable Compton angle.}
\end{figure}

A further ineluctable experimental effect, which impacts the performance of the Compton camera is the (noise-rejecting) threshold in the Scatter detector. Due to the Compton scattering law, a cut in deposited energy is directly translated into a threshold for the lowest measurable Compton angle. For $\gamma$-quanta of 662~keV this relationship is illustrated in Fig.~\ref{fig:threshold}. A threshold of 100~keV in the $S$-detector therefore leads to a bottom limit on the detectable angle of $\theta_{min} \sim 30^{\circ}$. Such threshold constraint is also indicated by dashed lines in Fig.~\ref{fig:2dresolution}.

In summary, due to the aforementioned geometry- and threshold effects, only the central region of Compton angles becomes experimentally accessible (see Fig.~\ref{fig:2dresolution}). Although this statement strictly holds only for a centered $\gamma$-ray source and the simple geometric model used, it serves to provide a realistic idea about some of the main constraints in the design of a Compton camera. For a 100~keV noise-rejecting threshold in the S-detector and the $a=2\times s$ configuration, the range of measurable angles is constrained to 30$^{\circ} \lesssim \theta \lesssim 80^{\circ}-100^{\circ}$ for 662~keV $\gamma$-rays depending on $d_f$. The next sections present the assembly and realization of a Compton imager built according to the concepts discussed above.

\section{Materials}\label{sec:materials}
\begin{figure}[htbp!]
\flushleft
\centering
\includegraphics[width=0.8\columnwidth]{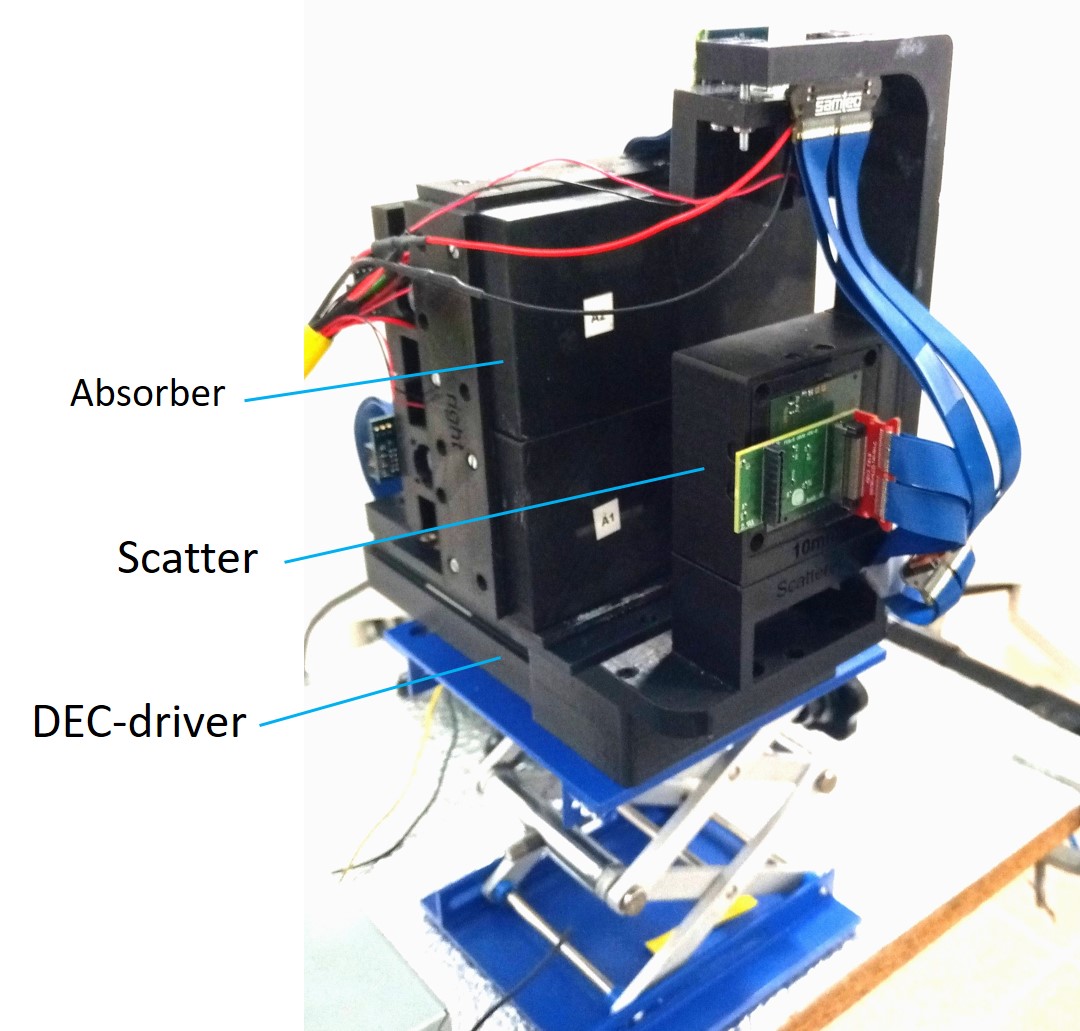}
\caption{\label{fig:i-TED} The i-TED Compton camera equipped with one scatter and four absorber detectors mounted on a positioning drive for the implementation of the DEC-technique. See also Fig.~\ref{fig:drive}}.
\end{figure}

i-TED is based on the Compton camera design with the aforementioned $a=2\times s$ configuration. The position-sensitive scatter detector (S) consists of a 50$\times$50$\times$10~mm$^3$ \lacls monolithic crystal optically coupled with silicon grease (Saint Gobain BC-630) to a silicon photomultiplier SiPM (SensL ArrayJ-60035-64P-PCB), which features a segmentation of 8$\times$8 pixels. The absorber detector is based on an array made from four co-planar \lacls blocks, each of them with a size of 50$\times$50$\times$25~mm$^3$ \lacls and readout by the same type of SiPM sensors. Thus, the full detection system comprises 320 channels, which are readout by means of front-end and processing PETsys FEB/D-v2 electronics~\cite{petsys}. Thermal stabilization of the ASICs is accomplished by means of 20$\times$20~mm$^2$ Peltier cells (FPH1-7106NC) thermally coupled to the external chip surface by means of non-silicone heat transfer compound (HTCP20S from Electrolube). In order to optimize cooling performance, the hot side of the Peltier cell is also thermally coupled to a small metallic heatsink assembled to a mini DC-axial fan (MC36358 from Multicomp), which helps to dissipate heat efficiently. Further details about the instrumentation related to the position sensitive detectors can be found in Ref.~\cite{Babiano19}. A picture of the whole assembled detection system is shown in Fig.~\ref{fig:i-TED}.

\begin{figure}[htbp!]
\flushleft
\centering
\includegraphics[width=\columnwidth]{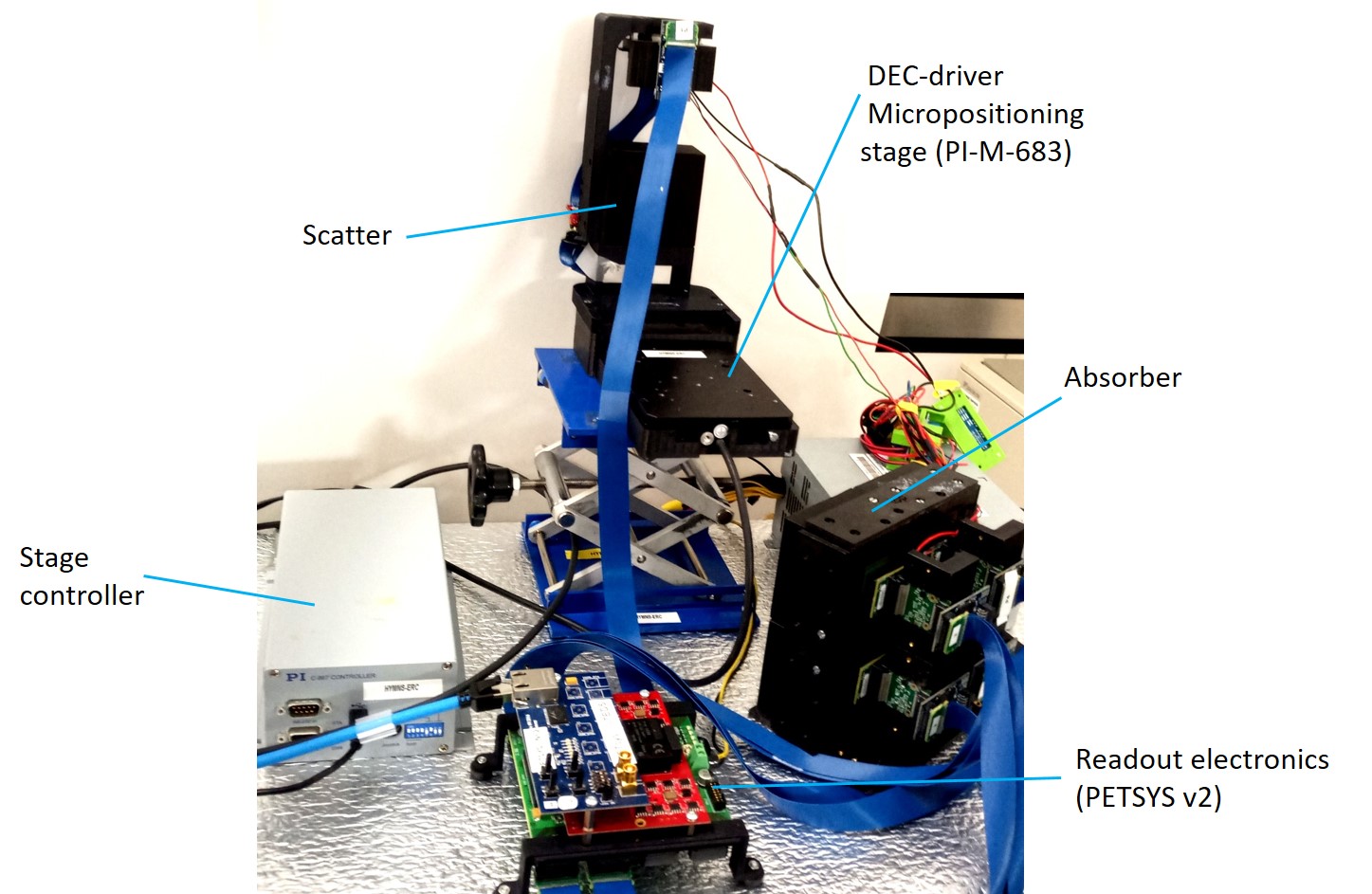}
\caption{\label{fig:drive} i-TED with A-layer taken off to show the positioning drive and other components.}
\end{figure}

Dynamic Electronic Collimation (see Sec.~\ref{sec:methods}) is accomplished in i-TED by means of a micropositioning stage (M-683 from PI-miCos) mounted underneath the array of A-detectors. This stage has a load capacity of 50~N and includes an integrated linear encoder with a resolution of 0.1~$\mu$m. The embedded piezoceramic linear motor (PILine U-164) allows one to remotely control the position of the absorption layer with respect to the scattering plane with a sub-micrometric precision over a range of 50~mm. Communication with the external computer is made by means of a controller (C-867 from PI-miCos) via USB connection. Fig.~\ref{fig:drive} shows a picture of the full set-up with the $A$-detection layer taken appart to illustrate better the different components.
The information about the distance $d_f$ between the $S$- and $A$-layers is directly fed into our data-stream and afterwards used for the image reconstruction (see Sec.~\ref{sec:results}).

\section{Measurements and results}\label{sec:results}

\begin{figure}[htbp!]
\flushleft
\centering
\includegraphics[width=0.95\columnwidth]{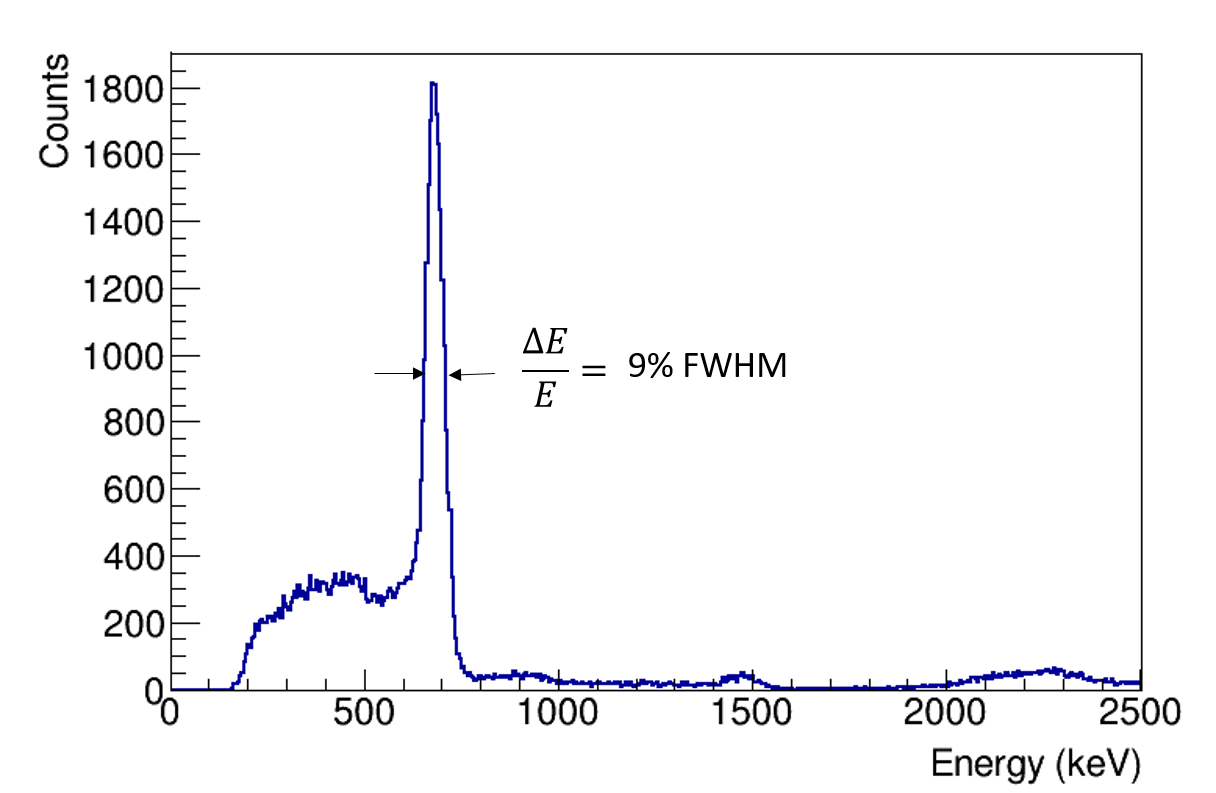}
\caption{\label{fig:Cs137} Sum spectrum for a $^{137}$Cs source with the $S$- and $A$-layers in time-coincidence.}
\end{figure}

An energy calibration of each detector is performed in the full energy range from 122~keV up to 1410~keV using a radioactive source of $^{152}$Eu. The energy resolution at 662~keV ranges between 6~\% and 10~\% at \textsc{fwhm} for all the crystals and the average value of the detection threshold is 100~keV. A time coincidence window of 10~ns is used between the S- and the A-detectors to select $\gamma$-ray hits corresponding to the same event. The add-back energy spectrum of the $S$ and $A$ layer in time coincidence is displayed in Fig.~\ref{fig:Cs137} and shows an average resolution of 9~\% \textsc{fwhm} at 662~keV.

The intrinsic position reconstruction in each crystal is made by fitting the analytic expression of Li et al.~\cite{Li10} as described in Ref.~\cite{Babiano19}. Thus, an intrinsic position resolution of about 1~mm \textsc{fwhm} is achieved for both S- and A-detectors, respectively. The depth-of-interaction (DOI) in each crystal is obtained from the measured area at half-height of the scintillation-photon distribution for each registered event. The present uncertainty on DOI is of about 5~mm, as described in Ref.~\cite{Babiano19}.

\subsection{Backprojection method}
\begin{figure}[htbp!]
\flushleft
\centering
\includegraphics[width=0.85\columnwidth]{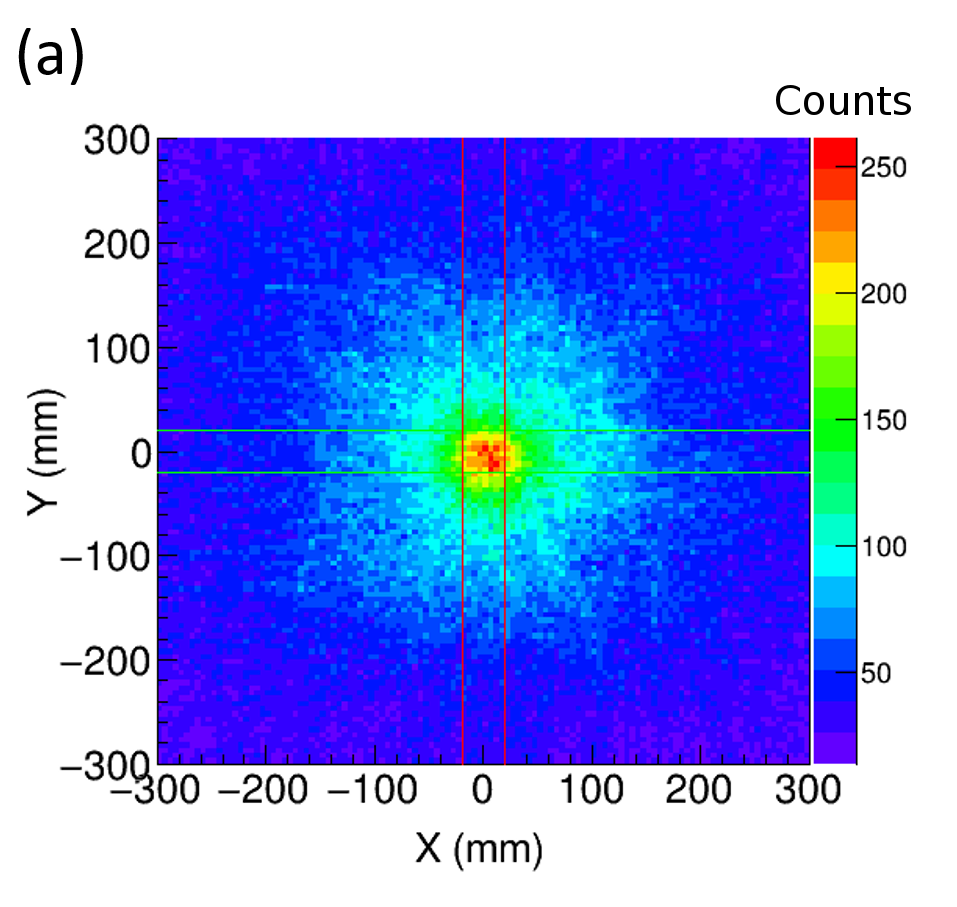}
\includegraphics[width=0.85\columnwidth]{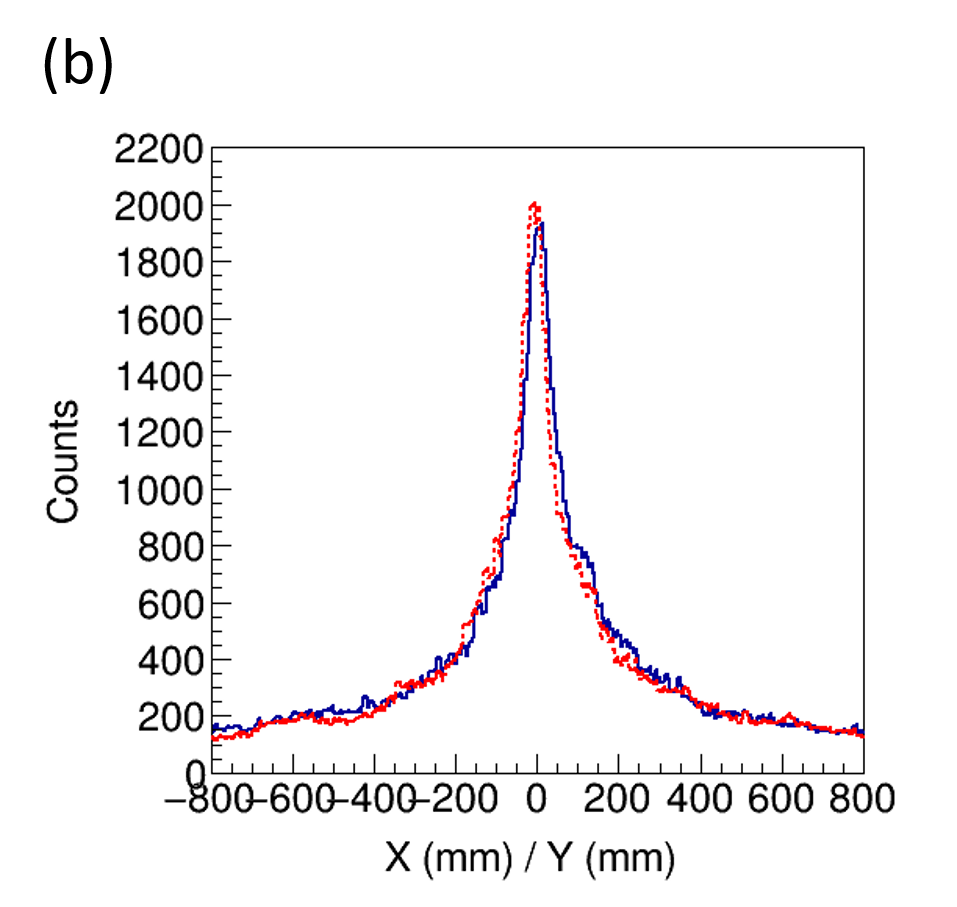}
\caption{\label{fig:BP center} Backprojected 2D-image (a) for a 511~keV $\gamma$-rays from a $^{22}$Na source centered at 165~mm in front of i-TED. Projections over the x- and y-axis are shown in panel (b).}
\end{figure}

A simple backprojection method has been implemented in order to reconstruct the Compton image at the plane of the $\gamma$-ray source. For illustrative purposes a point-like $^{22}$Na source with an activity of 416~kBq was centered at a distance of 165~mm in front of i-TED. The latter was operated with a $S$-$A$ separation length of $d_f = 30$~mm during a measuring time of 900~s. The image was formed by defining a pixelated plane at 165~mm in front of the detector with squared voxels of 5~mm size. The latter were then filled with the geometric intersection of each Compton cone with the plane, following a similar approach as the one reported in Ref.~\cite{Wilderman98}. The result obtained for the 2D-image is displayed in Fig.~\ref{fig:BP center}. A selection of $\pm$4~pixels, corresponding to $\pm$20~mm was made around the maximum of the reconstructed distribution in order to analyze the image projection over the $x$- and $y$-coordinates. The width of these projections is used below in Sec.~\ref{sec:resolution} as an estimate for the image resolution.

\subsection{Efficiency versus $S$-$A$ separation length $d_f$}\label{sec:efficiency}
\begin{figure}[htbp!]
\flushleft
\centering
\includegraphics[width=0.95\columnwidth]{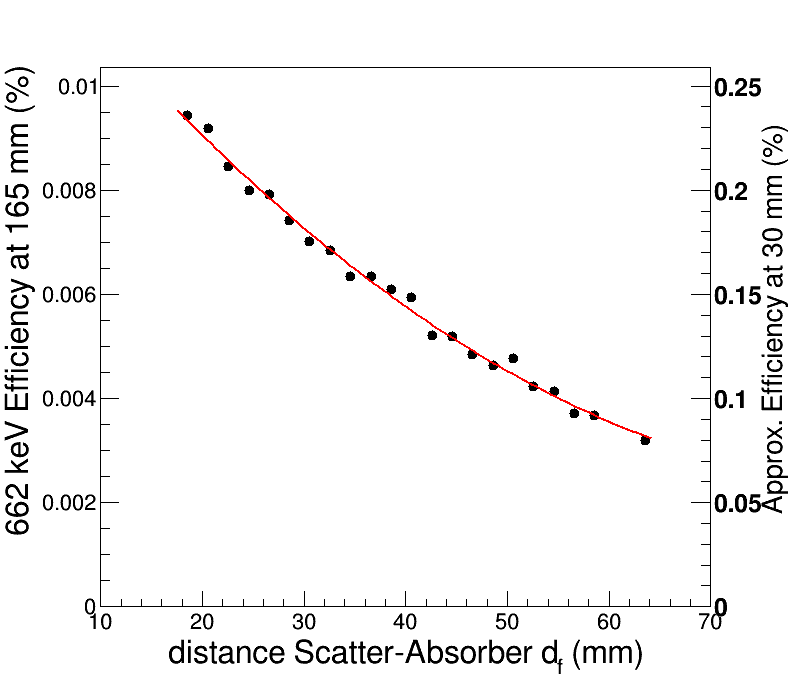}
\caption{\label{fig:efficiency} Measured efficiency for a point-like $^{137}$Cs source (662~keV) centered at 165~mm in front of i-TED. The right-axis shows the efficiency estimated for a distance of 30~mm.}
\end{figure}

The efficiency for 662~keV $\gamma$-rays was measured as a function of the distance between the $S$- and $A$-layers. To this aim a point-like $^{137}$Cs source with an activity of 210.4~kBq was placed at a distance of 165~mm from the center of the Compton camera. The efficiency was determined as a function of the distance between scatter and absorber detection planes by performing a series of measurements in steps of $\Delta d_f = 2$~mm. Each measurement lasted for 15 minutes. The results are displayed in Fig.~\ref{fig:efficiency} and show a dependency with $d_f$ (mm) given by $\varepsilon_{\gamma} = 1.358(14)\times10^{-2} -2.52(7)\times10^{-4} \cdot d_f + 1.41(8)\times10^{-6} \cdot d_f^2$ (\%). For comparative purposes the efficiency extrapolated at a distance of $d_s = 30$~mm is also shown on the right-hand side vertical axis. This will be used later in Sec.~\ref{sec:summary} for the discussion of the results. 
%

\subsection{Angular resolution versus $S$-$A$ separation length $d_f$}\label{sec:resolution}

\begin{figure}[htbp!]
\flushleft
\centering
\includegraphics[width=0.95\columnwidth]{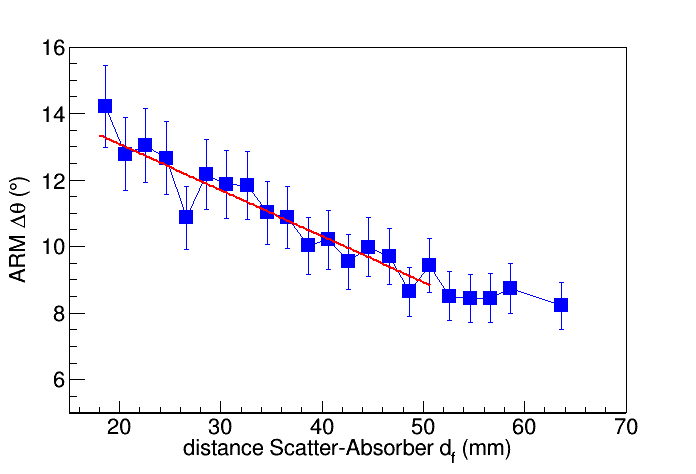}
\caption{\label{fig:resolution} Measured angular resolution as a function of the $S$-$A$-separation length $d_f$ using the backprojection method for a point-like $^{137}$Cs source (662~keV) placed at 165~mm in front of the center of i-TED.}
\end{figure}

The series of measurements described in the previous section were analyzed with the backprojection method and converted to angular units in order to determine the Angular Resolution Measure (ARM) to the level of one standard deviation as a function of the $S$-$A$ distance $d_f$. The results are displayed in Fig.~\ref{fig:resolution}.
The angular resolution shows an approximate linear behavior as a function of the distance $d_f$, with ARM values varying from about 14$^{\circ}$ to 9$^{\circ}$ for $d_f$ lengths of 20~mm and 50~mm, respectively. In this range, the ARM is described by $\Delta\theta = 15.8(9) - 0.14(2)\cdot d_{f}$ with $d_f$ in mm and ARM in degrees. Beyond $d_f = 50$~mm the angular resolution tends to a constant value of 8.5(3)$^{\circ}$. This trend is expected from the rather constant resolution shown in the diagrams of Fig.~\ref{fig:2dresolution} at large $d_f$ distances.

\subsection{Field of view}
\begin{figure}[htbp!]
\flushleft
\centering
\includegraphics[width=0.8\columnwidth]{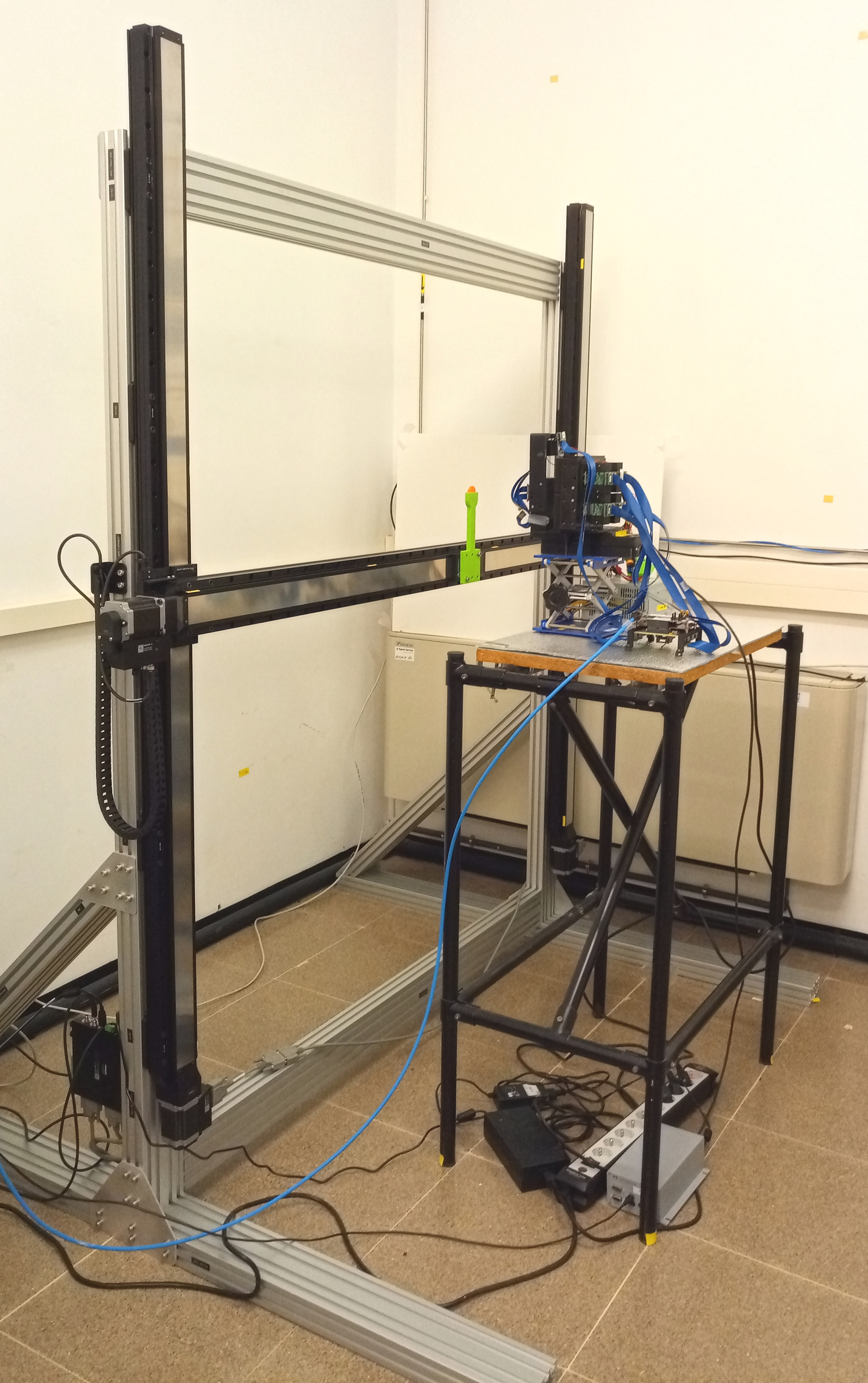}
\caption{\label{fig:gantry} Set-up used to evaluate the field-of-view, with i-TED in front of the vertical positioning gantry.}
\end{figure}

In order to explore the field of view of i-TED a measurement was carried out placing the $^{22}$Na source at nine different positions in front of the imager forming a cross with a spacing of 150~mm. To  prevent human error and eliminate any uncerainty related to the positioning of the source with respect to the imager a vertical gantry was set-up as shown in Fig.~\ref{fig:gantry} using LRT1500AL linear stages from Zaber Technologies Inc. The latter have an accuracy of 375~$\mu$m and a repeteability error of $<2$~$\mu$m.

\begin{figure}[!htbp]
\flushleft
\centering
\includegraphics[width=0.85\columnwidth]{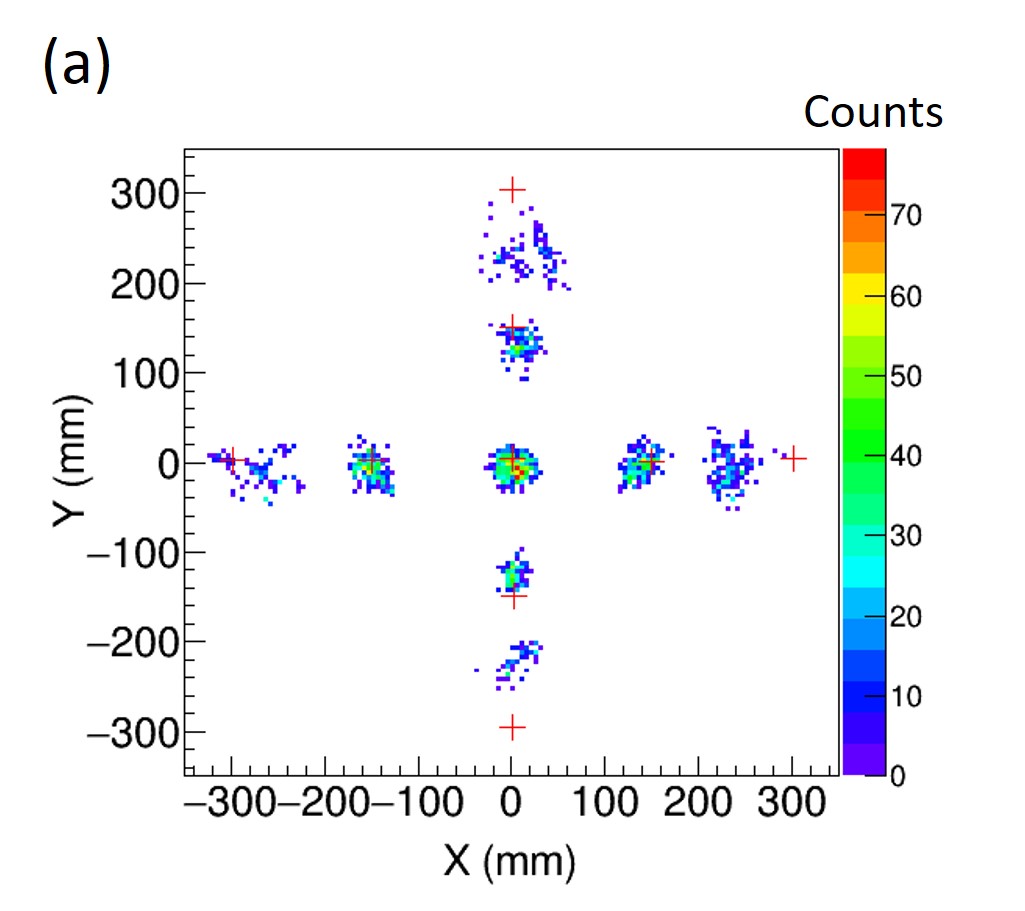}
\includegraphics[width=0.85\columnwidth]{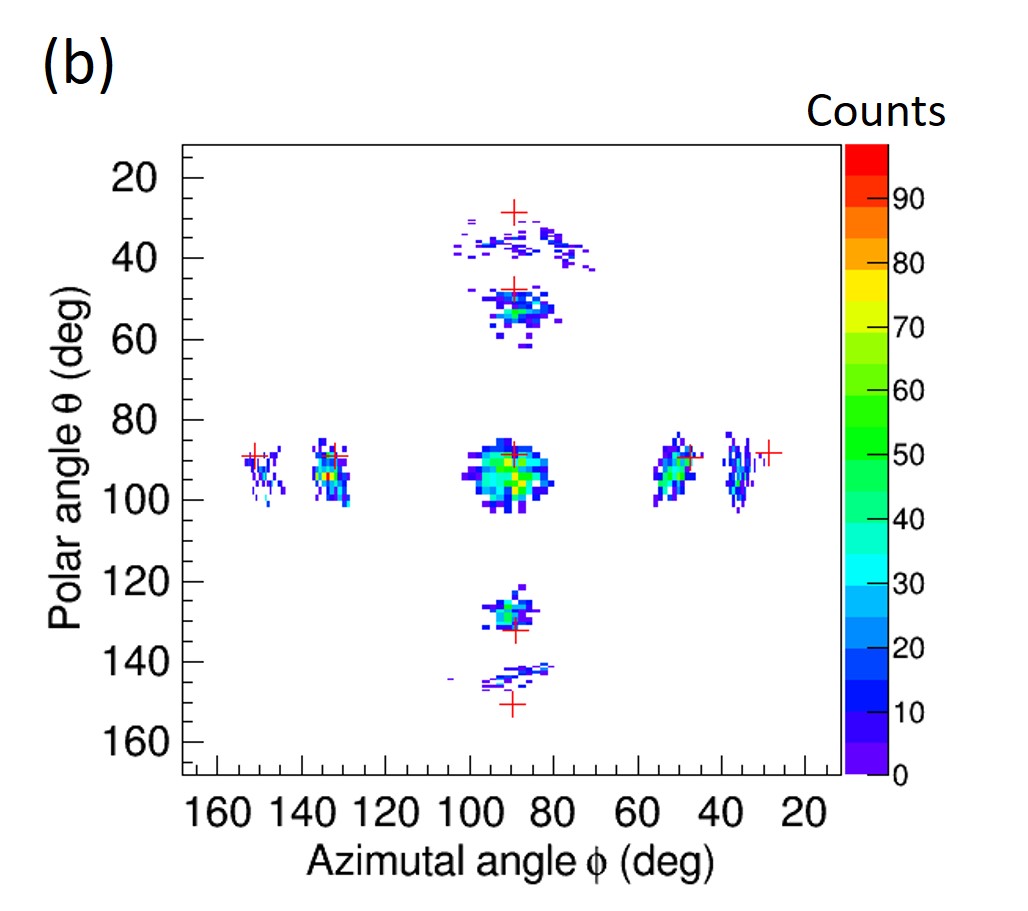}
\caption{\label{fig:BP} Positions reconstructed with the backprojection algorithm in Cartesian coordinates (a) and spherical coordinates (b). Small cross-marks indicate the true-positions of the source.}
\end{figure}

The gantry was controlled and synchronized with our data-acquisition system, in a similar way as described in Ref.~\cite{Babiano19}. The distance between the central position and the imager was of 165~mm. $S$-$A$ separation length of $d_f = 30$~mm was used in these measurements. The results for the reconstructed Compton images are shown in Fig.~\ref{fig:BP} both in Cartesian- and spherical coordinates.
In the latter representation the zenith direction or reference for the $\theta$-coordinate coincides with the vertical $y$ direction of the cartesian coordinates, whereas the azimuthal angle $\phi$ is given with respect to the $x$-axis. In these figures a constant level background was determined and subtracted for each of the measured positions. Considering the simple backprojection reconstruction method implemented here the results are rather satisfactory because the reconstructed source positions coincide reasonably well with the true source positions in the main part of the field of view. A certain compression and loss of sensitivity can be appreciated in the peripheral positions. This effect can be most probably ascribed to the higher contribution of the peripheral region of the $S$ detector, where the linerarity is not as good as in the central region~\cite{Babiano19}. Better performances could be probably obtained by means of a more sophisticated method such as maximum-likelihood estimation~\cite{Sarmouk94,Parra98,Xu07}. However, this is out of the scope of the present work. Still, an angular field of view of about 2/3 of 2$\pi$ can be estimated from this measurement using the backprojection approach.

\section{Discussion and outlook}\label{sec:summary}

i-TED is a high-efficiency Compton imager, which implements the new concept of Dynamic Electronic Collimation for enhanced fieldability and optimization of the measuring time. In this work an i-TED demonstrator based on five large \lacls monolithic crystals and 320 readout channels has been assembled and characterized in terms of efficiency and imaging capabilities.

Efficiency is a key aspect for i-TED in order to keep the total measuring time within reasonable limits in neutron capture TOF experiments. Compared to other scintillator-based Compton cameras with efficiencies reported in a similar energy range~\cite{Kishimoto17,Nagao18}, the efficiency found here is a factor of $\sim$50 higher. This feature can be ascribed to the large sensitive volume of i-TED and the use of large and continuous (non-pixelated) scintillation crystals.

Angular resolutions between 8$^{\circ}$ and 14$^{\circ}$ have been obtained so far, which are quite satisfactory when considering the simple backprojection algorithm implemented. Such angular sensitivity should enable a significant reduction of the $\gamma$-ray background associated with external background sources in neutron capture TOF experiments~\cite{Domingo16}. The angular resolution found here is comparable to values reported recently, for example in Ref.~\cite{Kataoka13,Nagao18} for a similar $S$-$A$ separation of $d_f \sim 20$~mm. On the other hand, our resolution is approximately a factor of $\sim$2 worse than the values reported in Ref.~\cite{Kishimoto17,Munoz17} for similar $d_f$ values. This difference can be ascribed to the maximum likelihood expectation maximization (ML-EM) method used in the latter work for the position reconstruction, and also to the better spectroscopic performance of their detectors.

There are several aspects related to the performance of i-TED which we plan to improve. One of them concerns the energy resolution of several of the detectors used, which was above the expected value due to a few channels missing in the readout electronics chain. An improved energy resolution will lead to a lower angular incertitude and improved overall imaging performance or, correspondingly, background rejection. Additionally, we plan to reduce the energy threshold of our $S$ detector, which in this work was at 100~keV. This should yield a better sensitivity to the Compton angles in the low range ($10^{\circ} \lesssim \theta \lesssim 30^{\circ}$) of the high-resolution angular regime (see Fig.\ref{fig:2dresolution}). The i-TED background rejection algorithm conceived thus far for neutron capture TOF experiments is based on an analytic (backprojection) event-by-event decision, as described in Ref.~\cite{Domingo16}. In this respect we plan to explore the viability to combine more sophisticated maximum-likelihood expectation algorithms with the PHWT for the determination of the cross section.
Finally, next steps include new measurements with i-TED using the pulsed neutron-beam of CNA-Seville~\cite{Macias20}, as a preliminary study before the commissioning at CERN n\_TOF in 2021 after the long shutdown LS2. For the measurements at CNA we plan to implement $^6$LiH neutron absorbers to reduce the intrinsic neutron sensitivity of i-TED, as discussed in Ref.~\cite{Domingo16}. By exposing the detector to the neutron beam we plan to perform a neutron sensitivity study similar to the one carried out in the past at FZK using C$_6$D$_6$ detectors~\cite{Plag03}.

\section*{Declaration of competing interest}
The authors declare that they have no known competing financial interests or personal relationships that could have appeared to influence the work reported in this paper.

\section*{CRediT authorship contribution statement}
\textbf{V. Babiano:} Software, Formal analysis, Investigation, Visualization. \textbf{J. Balibrea:} Visualization, Investigation, Writing -review \& editing. \textbf{L. Caballero:} Data curation, Investigation, Validation. \textbf{D. Calvo:} Resources, Investigation. \textbf{I. Ladarescu:} Software, Investigation, Formal analysis, Data curation. \textbf{J. Lerendegui:} Visualization, Investigation, Writing -review \& editing. \textbf{S. Mira Prats:} Methodology, Investigation. \textbf{C. Domingo-Pardo:} Conceptualization, Methodology, Supervision, Project administration, Writing - original draft, Writing -review \& editing, Funding acquisition, Software, Investigation.

\section*{Acknowledgment}
This work has received funding from the European Research Council (ERC) under the European Union's Horizon 2020 research and innovation programme (ERC Consolidator Grant project HYMNS, with grant agreement n$^{\circ}$ 681740). The authors acknowledge support from the Spanish Ministerio de Ciencia e Innovaci\'on under grants FPA2014-52823-C2-1-P, FPA2017-83946-C2-1-P, CSIC for funding PIE-201750I26 and the program Severo Ochoa (SEV-2014-0398). 

\bibliography{bibliography}

\end{document}